\documentstyle[12pt]{article}
\begin{document}
\begin{titlepage}
\title{
{\bf Dual Statistical Systems and Geometrical String\\
}
}
{\bf
\author{
 George K. Savvidy\\
 Physics Department, University of Crete\\
 71409 Iraklion, Crete, Greece\\
 and\\
 Yerevan Physics Institute, 375036 Yerevan, Armenia \vspace{1cm}\\
 Konstantin G.Savvidy\\
 Princeton University, Department of Physics\\
 P.O.Box 708, Princeton, New Jersey 08544 \vspace{1cm}\\
 Paul G.Savvidy\\
 Physics Department, University of Crete \\
 71409 Iraklion, Crete, Greece
}
}
\date{}
\maketitle
\begin{abstract}
\noindent

We analyse statistical system with interface energy
proportional to the length of the edges of interface.
We have found the dual system high temperature expansion
of which equally well generates surfaces with linear
amplitude. These dual systems are in the same relation
as 3D Ising ferromagnet to the 3D Gauge spin system.

\end{abstract}
\thispagestyle{empty}
\end{titlepage}
\pagestyle{empty}

{\Large{\bf 1.}}
Low temperature expansion of the Ising ferromagnet has
a beautiful geometrical representation
of the partition function in terms
of the sum over the paths on the
two-dimensional lattice and over the surfaces
on three-dimensional lattice. The corresponding amplitudes are
proportional to the length
of the paths and to the $area$ of the random surfaces.
This geometrical representation allows
to show, that  high temperature expansion of 2D Ising
ferromagnet is dual to that of low temperature expansion
and that the system is therefore self-dual
\cite{waerden,kramers,ising,onsager,kac,wegner}.
In the case of 3D Ising ferromagnet

$$H^{3D}_{Ising} =-\sum_{links}\sigma \sigma \eqno(1a)$$
the high temperature expansion
does not coincide with its low temperature
expansion and can be represented as a sum over the random
walks with the amplitude which is proportional to the $length$
of the paths.

In 1971 Franz Wegner succeeded to construct the spin system
with local
interaction low temperature expansion of which coincides
with the random
walks in 3D with the $length$ amplitude and with high
temperature
expansion of which coincides with the sum over the
random surfaces with the $area$ amplitude.
In this natural way he have found the Gauge spin Hamiltonian
which is dual to 3D Ising ferromagnet  \cite{wegner}

$$H^{3D}_{Gauge} = -\sum_{plaquettes}
\sigma \sigma \sigma \sigma . \eqno(1b)$$
One motivation for the study of the spin systems
(1) is that they can help to understand the dynamics
of random surfaces with area action
\cite{marinari,frohlich,orland,fradkin,itzykson,casher,polyakov}

In the recent articles  \cite{wegner1,savvidy}
the authors formulated a new class of statistical systems,
whose interface energy is associated with the edges of
the interface. These systems again have geometrical representation
of the low temperature partition function in terms of the
sum over paths and surfaces, but with amplitude which
is proportional
to total $\it curvature$ of the path on the two-dimensional
lattice and to the linear $\it size$ of the surface in three
dimensions
\cite{ambartzumian,savvidy1,savvidy2,schneider,durhuus,baillie}.

In terms of Ising spin variables $\sigma_{\vec r}$
the Hamiltonian
of this system has the form \cite{wegner1,savvidy}

$$ H^{3D}_{Gonihedric} = - \sum_{\vec r,\vec \alpha,\vec \beta}
\sigma_{\vec r} \sigma_{\vec r+\vec \alpha}
\sigma_{\vec r+\vec \alpha+\vec \beta}
\sigma_{\vec r+\vec \beta} \eqno(2).$$
where $\vec r$ is a three-dimensional vector on the lattice
$Z^{3}$ whose components are integer and $\vec \alpha$ ,$\vec \beta$
are unit vectors parallel to axis. Together with usual
symmetry group $Z_{2}$ the system (2) has an extra symmetry:
one can independently flip
the spins on any combination of planes (spin layers).
We should stress that the Ising spins in (2) are on the vertices
of the lattice $Z^{3}$ and are not on the links, as it is
in (1b).

It is an important fact that there exist dual
representation of the  random surfaces on the lattice
with area law in terms of Ising and
Wegner Hamiltonians (1a), (1b)
\cite{nielsen,gross,kazakov,ambjorn,david}.
Our aim is to construct
Hamiltonian which is dual to $H^{3D}_{Gonihedric}$ (2)
and equally well describes surfaces with linear amplitude
and to compare these complementary systems with each other.
\vspace{1cm}

{\Large{\bf2.}} The partition function of the system (2)
can be written in the form

$$  Z(\beta) = \sum_{\{\sigma\}}
exp\{-\beta  H^{3D}_{Gonihedric} \} \eqno(3) $$
and as it was shown in \cite{wegner1,savvidy} the
low temperature expansion is extended
over all closed surfaces $\{ M \}$
with the restriction that only an $\it even$ number of
plaquettes
can intersect at any given $edge$  ( $2r = 0,2,4.$) and that only
one plaquette is at every given place

$$ Z(\beta) = \sum_{\{ M \}}
exp\{- 2 \beta A(M) \} ,\eqno(4a) $$
where $A(M)$ is linear-gonihedric action
\cite{ambartzumian,savvidy1,savvidy2}
which is equal to

$$A(M) = \sum_{<i,j>}  \lambda_{i,j}
\cdot \vert \pi - \alpha_{i,j} \vert ,\eqno(4b) $$
the summation is over
all edges~ $<i,j>$~ of~ $M$,~$\alpha_{i,j}$~ is the angle
between two neighbour
plaquettes of $M$ in $Z^{3}$ having a common edge $<i,j>$
of the length~ $\lambda_{i,j}$. As it is easy to see
$A(M)$ is proportional to the linear size of the surface $M$
\cite{ambartzumian,savvidy1,savvidy2}.

The Hamiltonian (2) and the partition function (3) provides
the representation (4) of the randomly fluctuating
surfaces on the lattice with gonihedric action $A(M)$
in terms of locally interacting fields-spins
\cite{wegner1,savvidy}.
To construct the dual Hamiltonian we should find the
geometrical representation of the high temperature
expansion of the system (2),(3). Let us consider
for that high temperature expansion of (3)

$$ Z(\beta) = \sum_{\{ \sigma \}}
\prod_{plaquetts} ch \beta \cdot
\{ 1 + th \beta \cdot (\sigma \sigma
\sigma \sigma ) \}. \eqno(5)$$
Opening the brackets and summing over $\sigma$
one can see that only
those terms produce nonzero contribution which contain an
even number of plaquettes on every given $vertex$, therefore

$$Z(\beta) = (2 ch \beta )^{3N^{3}}
\sum_{\{\Sigma \}} (th\beta)^{s(\Sigma)}~~, \eqno(6)$$
where the summation is extended over
all surfaces $\{ \Sigma \}$
with an even number of plaquettes at any given vertex.
The $s(\Sigma)$ is the number of
plaquettes of $\Sigma$, e.g. the area of
the surface. Open surfaces are allowed.
In the next section we will describe in details this set of
surfaces $\{\Sigma \}$ and will introduce the concept of
the group structure on this set.
\vspace{1cm}

{\Large{\bf3.}} As we have seen, the high temperature
expansion (6) of the gonihedric system (2), (3) in 3D is
extended over surfaces
$\{ \Sigma \}$ which can be considered as
a collection of plaquettes on a
cubic lattice with the restriction that only an
even number of plaquettes can intersect
at any given $vertex$
and that only one plaquette is at every given place.

Let us attach plaquette variables $U_{P}$ to each plaquette
$P$ of $Z^{3}$

$$ U_{P} = -1~~~~ if~~ P \in~~ M ~~~~and~~~~U_{P}
= 1~~~~if~~\not\in M \eqno(7)$$
There are twelve plaquettes $P$ incident to every vertex
of the lattice. The constraint on the plaquette variables $U_{P}$
in every vertex

$$ \prod_{12~plaquettes~incident~to~vertex}
U_{P}=1, \eqno(8)$$
uniquely characterizes our set
of surfaces $\{ \Sigma \}$ in (6).

Now one can introduce the group structure on this set
of surfaces $\{ \Sigma \}$ (8).
Let us consider two surfaces $\Sigma^{1}$ and
$\Sigma^{2}$ and denote their plaquette variables
as $U^{1}_{P}$ and $U^{2}_{P}$ respectively.
Let us define the group product of those two surfaces as

$$U_{P} = U^{1}_{P} \cdot U^{2}_{P}. \eqno(9)$$
According to this definition a given plaquette
belongs to a group product of two surfaces
$\Sigma = \Sigma^{1} \otimes \Sigma^{2}$ only if it
belongs  exactly to one of $\Sigma^{1}$ or $\Sigma^{2}$.

One should check that the group product defined in
this way leaves the surfaces in the same class (8).
Indeed if (8) holds for $\Sigma^{1}$ and $\Sigma^{2}$
then it holds for the surface product $\Sigma$ ,
that is $\Sigma$ also have even number of plaquettes
on its every vertex (8). The inverse element of $\Sigma$
coincides with itself. The set of surfaces $\{ \Sigma \}$ (8)
finally forms an Abelian group $G$.

One can show that the whole group $G$ is a product
of the local group $G_{\xi}$. This group $G_{\xi}$
has four elements - elementary "matchbox" surfaces,

$$e = ~~~~~~~~~~,g_{\chi}=~~~~~~~~~~,g_{\eta}
= ~~~~~~~~~~,g_{\varsigma}= ~~~~~~~~~~ \eqno(10a)$$
with the multiplication table

$$e \cdot g_{\chi,\eta,\varsigma} = g_{\chi,\eta,\varsigma}, $$
$$g_{\chi} \cdot g_{\chi} = g_{\eta} \cdot g_{\eta} =
g_{\varsigma} \cdot g_{\varsigma} =e$$
$$g_{\chi} \cdot g_{\eta}= g_{\varsigma}, \eqno(11)$$
which follows from the multiplication law (9).

With the help of $G_{\xi}$ one can reconstruct any
surface $\Sigma$ of the set $\{ \Sigma \}$ (8) .
Indeed any set of elementary matchbox surfaces $e, g_{\chi},
g_{\eta}, g_{\varsigma}$ (10), (11) distributed independently
over the lattice $Z^{3}$ describes some
allowed surface $\Sigma$ and any given surface from
$\{ \Sigma \}$ (8) can be
decomposed into the product of $G_{\xi}$

$$\Sigma = \prod_{\xi} \cdot G_{\xi}. \eqno(12)$$

This approach allows to describe the original
surface $\Sigma$ in
terms of a new independent matchbox spin variable

$$G_{\xi}=\{ e(\xi), g_{\chi}(\xi),
g_{\eta}(\xi), g_{\varsigma }(\xi) \} ,\eqno(10b) $$
which should be attached to the centers of the
cubes $\xi$ of the original lattice
$Z^{3}$ or to vertices of the dual lattice
$Z^{\star~3}$.

This is what we aimed at: to express the surface configuration
$\Sigma$ with the constraints (8) in terms of independent local
variable $G_{\xi}$.
The group $G_{\xi}$ is an
Abelian group of the fourth
order and therefore has four one-dimensional
irreducible representations

$$E =\{1,1,1,1 \},~ R^{\chi}=
\{1,1,-1,-1 \}),~ R^{\eta}=\{1,-1,1,-1\},~R^{\varsigma}=
\{1,-1,-1,1\} \eqno(13a)$$
with the orthogonality relations

$$\sum_{G_{\xi}} R^{l}(G_{\xi}) \cdot R^{m}(G_{\xi})=
4 \delta^{l,m}~ (l,m = \chi , \eta ,\varsigma)$$
$$\sum_{G_{\xi}} R^{\chi}(G_{\xi})
R^{\eta}(G_{\xi}) R^{\varsigma}(G_{\xi}) = 4 \eqno(13b)$$
We will use this representations to express
algebraically the matchbox spin variable
$G_{\xi}$. The next step is to express the
amplitude of $\Sigma$
in terms of this independent variables and to
construct the dual Hamiltonian.
\vspace{1cm}

{\Large{\bf4.}}
In the high temperature expansion  (6),  the energy
of the surface $\Sigma$ is equal to the
number of plaquets $s(\Sigma)$, that is to area.
We would like to construct now a new
system of locally interacting
matchbox spins $G_{\xi}$ with identical
low temperature expansion. For that we should
properly organize local interaction
of the matchbox spins $G_{\xi}$.

The dual Hamiltonian is nonhomogeneous in the directions
$\chi$,~$\eta$,~and $\varsigma$ and is equal to

$$H_{dual} = \sum_{\xi} H_{\xi,\xi + \chi} +
H_{\xi,\xi +\eta} + H_{\xi,\xi + \varsigma}, \eqno(14)$$
where $\chi$,~$\eta$,~and $\varsigma$ are unit vectors
in the corresponding directions of the dual lattice and

$$H_{\xi,\xi + \chi} \equiv H(G_{\xi},G_{\xi + \chi})
= - R^{\chi}(\xi) \cdot R^{\chi}(\xi + \chi),$$
$$H_{\xi,\xi + \eta} \equiv H(G_{\xi},G_{\xi + \eta})
= - R^{\eta}(\xi) \cdot R^{\eta}(\xi + \eta),$$
$$H_{\xi,\xi + \varsigma} \equiv H(G_{\xi},G_{\xi + \varsigma})
= - R^{\varsigma}(\xi) \cdot R^{\varsigma}(\xi + \varsigma).
\eqno(15)$$
The partition function of the dual system (14), (15)
can be written in the form

$$  Z(\beta^{\star}) = \sum_{\{ G_{\xi} \}}
exp\{-\beta^{\star}  H_{dual} \}. \eqno(16)$$
Now we should check, that the low temperature expansion
of the dual system (16) indeed coincides with the
high temperature expansion of the original system (6).
As we will see this indeed takes place and we can
expect that high temperature expansion of the dual
Hamiltonian will provide us with the new lattice representation
of the random surfaces with gonihedric action.
In the next section we will show that two systems (2)
and (14), (15) are indeed dual to each other.
\vspace{1cm}

{\Large{\bf5.}}
Let us define the surface of interface $\Sigma$ for the
matchbox
spins $G_{\xi}$ in the following way: plaquette $P$ belongs
to $\Sigma$ if the product of neighboring matchbox spins is
equal to -1

$$P_{\xi,\xi +\chi} \in \Sigma ~~~if ~~~R^{\chi}(G_{\xi})
\cdot R^{\chi}(G_{\xi + \chi}) = -1 \eqno(17)$$
and

$$P_{\xi,\xi +\chi} \not\in \Sigma ~~~if ~~~R^{\chi}(G_{\xi})
\cdot R^{\chi}(G_{\xi + \chi}) = 1. \eqno(18)$$
In the same way we should define $P_{\xi,\xi +\eta}$ and
$P_{\xi,\xi +\varsigma}$.
These surfaces are of the
class $\{ \Sigma \}$ (8) because plaquette variables $U_{P}$
defined as

$$ U_{\xi,\xi +\chi} = R^{\chi}(G_{\xi}) \cdot
R^{\chi}(G_{\xi + \chi})$$
$$ U_{\xi,\xi +\eta} = R^{\eta}(G_{\xi}) \cdot
R^{\eta}(G_{\xi + \eta})$$
$$ U_{\xi,\xi +\varsigma} = R^{\varsigma}(G_{\xi}) \cdot
R^{\varsigma}(G_{\xi + \varsigma}) \eqno(19)$$
identically resolves the constranes (8).
The correspondence between matchbox
spin configurations and surfaces $\Sigma$ is four-to-one,
instead of two-to-one in the case of surfaces of interface of
the Ising ferromagnet.

By construction the energy of the
surface $\Sigma$ of matchbox
spin interface is equal to its area and therefore the
low temperature expansion
of the dual system (16) indeed coincides with the
high temperature expansion of the original system (6).

\vspace{1cm}

{\Large{\bf6.}} As the last step in this construction we
should prove that the high temperature expansion
of the dual system (14), (15) is equivalent to the sum
over  random
surfaces with the amplitude which is proportional
to the linear size of the surface  $A(M)$ and coincides
with the low temperature expansion (4 ) of (3).

Expanding the partition function (16) for small $\beta^{\star}$

$$  Z(\beta^{\star}) = \sum_{\{ G_{\xi} \}}
\prod_{\xi} (ch\beta^{\star})^{3}
(1 - th\beta^{\star}\cdot  H_{\xi,\xi + \chi})
(1 - th\beta^{\star} \cdot H_{\xi,\xi +\eta})
(1 - th\beta^{\star} \cdot H_{\xi,\xi + \varsigma})
\eqno(20)$$
and opening the brackets we will have the terms of the form

$$\sum_{\{ G_{\xi} \}} \prod_{over~paths~on~Z^{\star~3}}
R(\xi) \eqno(21) $$
representing the chains of $R's$ (13) of different lengths on
the lattice $Z^{\star~3}$. Let us see which of those terms
are nonzero. Using relations

$$ R^{\chi}(G_{\xi})R^{\chi}(G_{\xi})=
R^{\eta}(G_{\xi})R^{\eta}(G_{\xi}) =
R^{\varsigma}(G_{\xi})R^{\varsigma}(G_{\xi}) =1$$
$$R^{\chi}(G_{\xi})
R^{\eta}(G_{\xi}) R^{\varsigma}(G_{\xi}) = 1 \eqno(22)$$
which hold for the irreducible representations (13), one can
get

$$\sum_{G_{\xi}} H_{\xi - \chi, \xi} H_{\xi, \xi + \chi}
= 4 R^{\chi}(\xi -\chi)\cdot
R^{\chi}(\xi + \chi),$$
$$\sum_{G_{\xi}}
H_{\xi - \chi, \xi} H_{\xi, \xi + \eta}
=\sum_{G_{\xi}} H_{\xi -\chi, \xi + \varsigma}
= 0. \eqno(23)$$
Last two relations tell us that any loop which contains
one turn at the right angle is equal to zero. We have
three extra nonzero elementary verteces

$$\sum_{G_{\xi}} H_{\xi,\xi +\chi} H_{\xi, \xi + \eta}
H_{\xi, \xi + \varsigma} = -4
R^{\chi}(\xi +\chi)R^{\eta}(\xi + \eta)
R^{\varsigma}(\xi + \varsigma)$$
$$\sum_{G_{\xi}} H_{\xi -\chi,\xi} H_{\xi, \xi + \chi}
H_{\xi -\eta, \xi} H_{\xi, \xi +\eta} =
4 R^{\chi}(\xi -\chi)R^{\chi}(\xi + \chi)
R^{\eta}(\xi - \eta)R^{\eta}(\xi + \eta) $$
$$\sum_{G_{\xi}} H_{\xi -\chi,\xi} H_{\xi, \xi + \chi}
H_{\xi -\eta, \xi} H_{\xi, \xi +\eta}
H_{\xi -\varsigma, \xi} H_{\xi, \xi +\varsigma}=$$
$$4 R^{\chi}(\xi -\chi)R^{\chi}(\xi + \chi)
R^{\eta}(\xi - \eta)R^{\eta}(\xi + \eta)
R^{\varsigma}(\xi - \varsigma)
R^{\varsigma}(\xi + \varsigma)  \eqno(24)$$
and all other verteces are equal to zero.

So, high temperature expansion of the dual system (14), (15)
contains only those loops on the lattice $Z^{\star~3}$
which have only one of those four nonzero verteces of the type
(23) and (24). One can imagine every term of this expansion
as a "skeleton"  constructed by the loops of
"bones" restricted by the constraints (23) and (24).  The
amplitude is proportional to the total length of this
bones. The partition function therefore have the form

$$  Z(\beta^{\star}) = \sum_{skeletons}
(th\beta^{\star})^{A(skeleton)} \eqno(25)$$
where $A(skeleton)$ is the total length of the bones.

Nontrivial fact consists in the statement that one can
"dress up"  this skeletons by plaquettes so that
bones will appear as the right angle edges of the
surface $M$. In that case the $A(skeleton)$ becomes
identically equal to $A(M)$ and summation over
skeletons reduce to the summation over surfaces
$M$ with linear-gonihedric action $A(M)$. This
identification is possible only because  bones can
join together and form the loops only
though the vertices (23) and (24). So we have

$$ ln(th\beta^{\star}) = - 2 \beta \eqno(26)$$
and both systems (2) and (13),(14) and (15) are
dual to each other in the same way as 3D Ising
ferromagnet (1a) is dual to 3D Wegner gauge
Hamiltonian (1b). They are complementary to
each other in the sense that (1b) describe
random surfaces with area law and (14),(15)
with linear-gonihedric law.

Here raises the question of whether or not
the system exhibit a phase transition.
Because the energy functional is
proportional to the linear size of the surfaces
one can expect that the system will show phase
transition in 3D which should be of the same nature
as it is in the case of 2D Ising ferromagnet.

We gratefully acknowledge  stimulating discussions
with Franz Wegner. It is a pleasure to thank
J.Ambjorn, D.Gross, B.Durhuus, E.Marinari, H.Nielsen,
E.Paschos, E.Floratos, R.Flume, R.Schneider for
helpful conversations.

This work was sponsored in part by the Danish Natural
Science Research Council.

\vfill
\end{document}